# LIQUID-LIQUID EQUILIBRIA FOR SYSTEMS CONTAINING 4-PHENYLBUTAN-2-ONE OR BENZYL ETHANOATE AND SELECTED ALKANES

CRISTINA ALONSO TRISTÁN[(1)], JUAN ANTONIO GONZÁLEZ*[(2)], FERNANDO HEVIA,[(2)], ISAÍAS GARCÍA DE LA FUENTE[(2)] AND JOSÉ CARLOS COBOS[(2)]

[(1)] Dpto. Ingeniería Electromecánica. Escuela Politécnica Superior. Avda. Cantabria s/n. 09006 Burgos, (Spain)

[(2)] G.E.T.E.F., Departamento de Física Aplicada, Facultad de Ciencias, Universidad de Valladolid, Paseo de Belén, 7, 47011 Valladolid, Spain,

*e-mail: jagl@termo.uva.es; Fax: +34-983-423136; Tel: +34-983-423757

**ABSTRACT**

The method of the critical opalescence with a laser scattering technique has been employed for the determination of the liquid-liquid equilibrium (LLE) curves for the systems 4-phenylbutan-2-one + $CH_3(CH_2)_nCH_3$ ($n$ = 10,12,14) and for benzyl ethanoate + $CH_3(CH_2)_nCH_3$ ($n$ = 12,14). The mixtures are characterized by having an upper critical solution temperature (UCST), which increases with $n$. The corresponding LLE curves have a rather horizontal top and become shifted to higher concentration of the polar compound when $n$ is increased. Calorimetric data and LLE measurements show that the aromaticity effect leads to stronger interactions between molecules of the polar compound considered with respect to those between homomorphic linear molecules with the same functional group. This has been ascribed to proximity effects arising from the presence of the polar group and the aromatic ring within the same molecule. Proximity effects become weaker in the sequence: 1-phenylpropan-2-one > 4-phenylbutan-2-one > 1-phenylethanone, and are more relevant in benzyl ethanoate than in ethyl benzoate molecules. The DISQUAC group contribution model represents correctly the coordinates of the critical points of the investigated systems and the symmetry of the LLE curves.

## 1. Introduction

This work is part of a series concerned with the research of mixtures involving aromatic polar compounds. Up to now, we have investigated systems with aromatic amines[1-8] [1-8] (anilines, 2-amino-1-methylbenzene, 1-phenylmethanamine, 1$H$-pyrrole, quinoline or imizadoles); aromatic alcohols[9-10] (phenol or phenylmethanol), or aromatic alkanals, ketones or alkanoates.[11-13] Now, we report LLE measurements for 4-phenylbutan-2-one or benzyl ethanoate + alkane systems. Previously, we have provided LLE data for 4-phenylbutan-2-one + decane, or benzyl ethanoate + dodecane mixtures.[13] The presence of the phenyl group and of a polar group X in the same molecule (in this investigation, X = CO, or COO) leads to the existence of proximity effects between the mentioned groups. Such effects are of intramolecular character and it is well known that may have a decisive influence on the interaction parameters of the statistical model applied for the characterization of the systems under study. For example, main groups for phenol or aniline have been defined in the framework of the UNIFAC model (Dortmund version)[14] for a better representation of the thermodynamic properties of their mixtures. The data reported in this work are used to improve the DISQUAC[15,16] matrix of interaction parameters for contacts where the CO or COO groups participate.[11,13] On the other hand, the investigation of systems with aromatic heteroatoms is needed for a deeper understanding of the $\pi$-$\pi$ interactions and of the non-conventional H-bonds.[17,18] The functional CO and COO groups are particularly relevant. The former is encountered in proteins and hormones. The COO group attached to an aromatic ring is commonly used as amine protecting group of amino acids and it is of importance in peptide synthesis.[19] Aromatic esters are widely used in the manufacture of odorants and flavouring chemicals.

## 2. Experimental

*2.1 Materials.* Table 1 shows some properties of the pure compounds used along the work: source, purity, water contents, determined by the Karl-Fischer method, and density ($\rho$). The chemicals were employed without further purification. Densities were determined using a vibrating-tube densimeter and a sound analyser, Anton Paar model DSA-5000. The repeatability of the $\rho$ measurements is $5 \cdot 10^{-3}$ kg·m$^{-3}$, while their relative standard uncertainty is 0.0001. Accordingly to the results shown in Table 1, there is a good agreement between our $\rho$ measured values and literature data.

*2.2 Apparatus and Procedure*

Mixtures were prepared by mass by means of an analytical balance HR-202 (weighing accuracy $10^{-5}$ g), in small Pyrex tubes (0.009 m i.d. and about 0.04 m length). The tubes were immediately sealed by capping at 0.1 MPa and 298.15 K. Mole fractions were calculated using the relative atomic mass table of 2015 issued by the Commission on Isotopic Abundances and Atomic Weights (IUPAC)[20]

As in previous applications, the coexistence curves of liquid-liquid equilibrium were determined by the method of the critical opalescence. Details of the experimental method can be found elsewhere.[21] The equilibrium temperatures were measured by means of a Pt-1000 resistance, calibrated according to the ITS-90 scale of temperature, against the triple point of the water and the fusion point of Ga. The precision of the equilibrium temperature measurements is 0.001 K, with an estimated standard uncertainty of 0.05 K. The reproducibility of the mentioned temperatures is 0.02 K for values close to the upper critical solution temperature (UCST). For the equilibrium composition measurements, the standard uncertainty of the mole fraction is 0.0005. This estimation takes also into account that the more volatile compound is partially evaporated to the free volume of the ampoule ($\approx 1.17 \cdot 10^{-6}$ m$^3$).

### 3. Experimental results

Table 2 lists the directly measured liquid-liquid equilibrium temperatures, $T$, vs. the mole fraction of the aromatic polar compound, $x_1$, for the mixtures: 4-phenylbutan-2-one + dodecane, or + tetradecane, or + hexadecane, or benzyl ethanoate + tetradecane, or + hexadecane. No data have been encountered in the literature for comparison. The data are shown graphically in Figures 1 and 2.

The coexistence curves determined show a rather flat maximum and become progressively skewed towards higher $x_1$ values when the alkane size is increased in mixtures with a given polar component. In addition, the UCST of these systems increases with the alkane size. These features are also encountered in many others mixtures previously investigated as those formed by alkane and 1-phenylethanone,[11] or 1-phenylpropan-2-one[13] or phenylmethanal,[12] or aromatic alcohols,[10] or aromatic amines, or linear organic carbonate, or acetic anhydride, or alkoxyethanol, or linear polyether (source of experimental data may be found in reference[13]).

As usually, the experimental ($x_1$, $T$) pairs obtained for each system were correlated with the equation:[22,23]

$$T/K = T_c/K + k|y - y_c|^m \tag{1}$$

where

$$y = \frac{\alpha x_1}{1 + x_1(\alpha - 1)} \tag{2}$$

$$y_c = \frac{\alpha \, x_{1c}}{1 + x_{1c}(\alpha - 1)} \qquad (3)$$

In equations (1)-(3), $m$, $k$, $\alpha$, $T_c$ and $x_{1c}$ are the parameters which must be fitted against the experimental data. Particularly, ($x_{1c}$, $T_c$) stand for the coordinates of the critical point. It should be remarked that, when $\alpha = 1$, equation (1) is similar to :[24-26]

$$\Delta \lambda = B \tau^{\beta} \qquad (4)$$

where $\Delta \lambda_1 = \lambda_1' - \lambda_2''$ is the so-called order parameter, $\tau$ ($= T_c - T)/T_c$ ) represents the reduced temperature and $\beta$ denotes the critical exponent connected to $\Delta \lambda_1$. The order parameter may be any density variable in the conjugate phase (here, $\lambda_1 = x_1$). The value of the critical exponent $\beta$ depends on the theory applied to its determination.[24,27]

The adjustment of the $m$, $k$, $\alpha$, $T_c$ and $x_{1c}$ parameters was conducted by means of the Marquardt algorithm[28] with all the points weighted equally. Values of the fitted parameters and of the standard deviations for LLE temperatures, $\sigma(T)$, are given in Table 3, The standard deviations are calculated from:

$$(\sigma(T)/K) = \left[ \sum (T_{exp} - T_{calc})^2 / (N - n) \right]^{1/2} \qquad (5)$$

Here, $N$ and $n$ are, respectively, the number of data points and the number of fitted parameters. Equation (1) fits well the experimental measurements.

4.  **Discussion**

Along the present section, we are referring to $H_m^E$ values at 298.15 K and $x_1 = 0.5$. Firstly, we note that the $H_m^E$ values of alkane systems including an aromatic polar compound are larger than those of mixtures where the polar group is situated within a linear chain. For example, $H_m^E (n\text{-}C_7)/\text{J·mol}^{-1}$ = 1492 (1-phenylethanone);[29] 886 (heptan-2-one);[30] 1361 (phenylmethanal);[31] 1066 (pentanal);[32] 1154 (ethyl benzoate);[33] 528 (ethyl hexanoate).[34] Therefore, it is possible to conclude that interactions between like molecules are stronger in mixtures including aromatic polar compounds. The existence of miscibility gaps for such systems at temperatures not far from 298.15 K confirms this point (Figure 3).[11,13] The observed behaviour can be ascribed to the existence of proximity effects (intramolecular effects) between the aromatic ring and the polar group within the same molecule. In contrast, intermolecular effects are present in systems of the type linear polar compound + benzene, where the aromatic ring and polar group belong to different molecules. These intermolecular effects lead to lower $H_m^E$ values than those given above for aromatic polar compound + heptane mixtures, as it is indicated by the following examples: $H_m^E$(benzene)/J·mol$^{-1}$ = 138 (2-propanone);[35] −171 (2-hexanone);[36] 54 (propanal); −82 (pentanal);[37] 84 (ethyl ethanoate).[33]

The variation of proximity effects with the distance between the polar group and the aromatic ring strongly depends on the polar group under consideration. In the case of aromatic alkanones, the mentioned effects become weaker in the sequence: 1-phenylpropan-2-one > 4-phenylbutan-2-one > 1-phenylethanone, as for mixtures with a given alkane, the corresponding values of UCST (Figure 3) and of $H_m^E$ decrease in the same order. Thus, $H_m^E$(n-C$_7$)/J·mol$^{-1}$ = 1680 (1-phenylpropan-2-one)[29] > 1604 (4-phenylbutan-2-one)[38] > 1480 (1-phenylethanone).[29] All this allows conclude that alkanone-alkanone interactions become weakened in the same sequence. Interactions between like molecules are also stronger in benzyl ethanoate systems than in those with ethyl benzoate, as $H_m^E$(n-C$_7$)/J·mol$^{-1}$ = 1783 (benzyl ethanoate);[39] 1154 (ethyl benzoate).[33] An inversion of this behaviour is encountered in mixtures containing heptane and phenol (UCST = 327.3 K)[40] or phenylmethanol (323.7 K).[10]

It is interesting to compare results for mixtures with isomeric polar molecules and an alkane. Here, we remark the rather large difference between the UCST values of heptane mixtures involving methyl 2-phenylacetate (278.7 K) or benzyl acetate (263.0 K),[41] what clearly indicates that interactions between like molecules are stronger in systems with the former aromatic ester. On the other hand, Figure 3 shows that the UCST values of 1-phenylpropan-2-one mixtures are higher than those of benzyl acetate systems. This reveals that interactions between like molecules are weaker in the latter mixtures. However, the $H_m^E$(n-C$_7$)/J·mol$^{-1}$ values change in the opposite way: 1783 (benzyl ethanoate)[39] > 1680 (1-phenylpropan-2-one),[29] what suggests that dispersive interactions are more relevant in mixtures with the ester.

In previous works,[11,13] we have characterized aromatic alkanone or alkanoate + alkane systems in terms of DISQUAC,[15,16] providing the needed interaction parameters. Here, we have explored the validity of the mentioned parameters for the representation of the LLE curves of the studied mixtures. Details on DISQUAC, equations and fitting procedure, can be found elsewhere.[42,43] We merely remark that the temperature dependence of the interaction parameters is expressed in terms of the DIS (dispersive) and QUAC (quasichemical) interchange coefficients,[42] $C_{st,l}^{DIS}$; $C_{st,l}^{QUAC}$ where s ≠ t are two contact surfaces present in the mixture and $l$ = 1 (Gibbs energy); $l$ = 2 (enthalpy); $l$ = 3 (heat capacity). The investigated systems are built by three surfaces: type a, aliphatic (CH$_3$, CH$_2$, in alkanes or aromatic polar compounds), type b, aromatic (C$_6$H$_5$ in aromatic polar compound) and type k (CO in aromatic ketone, or COO in aromatic ester). The three surfaces generate three contacts: (a,b); (a,k) and (b,k). The interchange coefficients of the (a,b) contacts are merely dispersive and are available in the literature.[44] The (b,k) contacts in aromatic alkanoate systems are also represented by DIS parameters only, while the remainder contacts are described by DIS and QUAC interchange coefficients.[13] The $C_{\text{bk},1}^{\text{DIS/QUAC}}$ (l =1,2,3) coefficients have been used without modification.[13] However, as in other previous DISQUAC studies as those on mixtures including *N,N*-dialkylamides,[42] or pyridine[45] or phenylmethanol,[10] or 1-phenylethanone,[11] calculations show that the $C_{\text{ak},1}^{\text{DIS}}$ coefficients must be assumed to be dependent on the alkane size (Table 4) in order to provide a correct representation or the coordinates of the critical points (Table 3). This is due to theoretical LLE calculations are developed assuming that $G_{\text{m}}^{\text{E}}$ (molar excess Gibbs energy) is an analytical function close to the critical point, as DISQUAC is a mean field theory. However, it is well known that thermodynamic functions are expressed in terms of scaling laws with universal critical exponents and universal scaling functions.[24] This also leads to the calculated LLE curves are more rounded than the experimental ones at temperatures near to the UCST (Figures 1,2) and too high at the UCST and too low at the LCST[24] (lower critical solution temperature). Nevertheless, an important result is that DISQUAC represents the change in the symmetry of the LLE curves for the benzyl ethanoate mixtures when the alkane size is increased (Figure 2).

## 5. Conclusions

LLE curves have been experimentally obtained for the mixtures 4-phenylbutan-2-one + dodecane, or + tetradecane, or + hexadecane and benzyl ethanoate + tetradecane, or + hexadecane. All the curves show an UCST, which increases with the alkane size. Alkanone-alkanone interactions become weaker in the order: 1-phenylpropan-2-one > 4-phenylbutan-2-one > 1-phenylethanone and proximity effects between the phenyl ring and the CO group become also weaker in the same sequence. Proximity effects are also more relevant in benzyl ethanoate than in ethyl benzoate. DISQUAC describes correctly the coordinates of the critical points and the shape of the LLE curves using dispersive Gibbs energy parameters dependent on the alkane for the $CH_2$/X (X =CO, or COO) contacts.

**Funding**

F. Hevia gratefully acknowledges the grant received from the program 'Ayudas para la Formación de Profesorado Universitario (convocatoria 2014), de los subprogramas de Formación y de Movilidad incluidos en el Programa Estatal de Promoción del Talento y su Empleabilidad, en el marco del Plan Estatal de Investigación Científica y Técnica y de Innovación 2013-2016, de la Secretaría de Estado de Educación, Formación Profesional y Universidades, Ministerio de Educación, Cultura y Deporte, Gobierno de España'.

**Table1**

**Properties of Pure Compounds at 0.1 MPa and 298.15 K[a]**

| Compound | CAS | Source | Initial mole fraction | $\rho$ [a]/kg·m$^{-3}$ Exp. | $\rho$ [a]/kg·m$^{-3}$ Lit. | water[b] content |
|---|---|---|---|---|---|---|
| 4-phenylbutan-2-one | 2550-26-7 | Sigma-Aldrich | ≥98% | 984.742 | 982.95[45] <br> 984.34[38] | 700 |
| Benzyl ethanoate | 140-11-4 | Sigma-Aldrich | ≥99% | 1051.10 | 1051.13[46] | 350 |
| Dodecane | 112-40-3 | Fluka | ≥98% | 745.508 | 745.32[47] | 25 |
| Tetradecane | 629-59-4 | Fluka | ≥99% | 759.275 | 759.316[47] | 25 |
| Hexadecane | 544-76-3 | Fluka | ≥99% | 770.221 | 770.316[47] | 33 |

[a]standard uncertainties are: $u(T) = 0.01$ K; $u(P) = 0.1$ kPa; the relative standard uncertainty for density is $u(\rho) = 0.0001$ and 0.02 for water content; [b]in mass fraction (ppm)

**Table 2**

**Experimental Liquid-Liquid Equilibrium Temperatures for Aromatic Polar Compound(1) + *n*-Alkane(2) Mixtures[a] at 0.1 MPa.**

| $x_1$ | T/K | $x_1$ | T/K |
|---|---|---|---|
| 4-Phenylbutan-2-one (1) + dodecane(2) | | | |
| 0.2232 | 279.26 | 0.5673 | 292.37 |
| 0.2653 | 285.20 | 0.5993 | 292.30 |
| 0.2833 | 286.39 | 0.6316 | 292.18 |
| 0.2929 | 287.07 | 0.6521 | 292.23 |
| 0.3250 | 288.89 | 0.6695 | 291.92 |
| 0.3431 | 289.62 | 0.6939 | 291.61 |
| 0.3688 | 290.62 | 0.7211 | 291.02 |
| 0.4059 | 291.42 | 0.7534 | 289.94 |
| 0.4605 | 292.30 | 0.7684 | 289.13 |
| 0.4775 | 292.42 | 0.7997 | 286.82 |
| 0.4844 | 292.52 | 0.8176 | 285.53 |
| 0.4853 | 292.48 | 0.8341 | 283.22 |
| 0.5528 | 292.38 | 0.8683 | 278.30 |
| 4-Phenylbutan-2-one (1) + tetradecane(2) | | | |
| 0.2262 | 285.78 | 0.5949 | 300.04 |
| 0.2421 | 287.33 | 0.6374 | 299.93 |
| 0.2911 | 291.60 | 0.6873 | 299.80 |
| 0.3173 | 293.39 | 0.7021 | 299.73 |
| 0.3346 | 294.67 | 0.7183 | 299.45 |
| 0.3524 | 295.47 | 0.7218 | 299.40 |
| 0.3731 | 296.42 | 0.7475 | 298.94 |

Table 2 (continued)

| | | | |
|---|---|---|---|
| 0.3958 | 297.41 | 0.7740 | 297.14 |
| 0.4479 | 298.98 | 0.7958 | 295.59 |
| 0.4610 | 299.26 | 0.8233 | 293.57 |
| 0.5040 | 299.96 | 0.8566 | 289.89 |
| 0.5259 | 300.13 | 0.8629 | 289.03 |
| 0.5511 | 300.08 | 0.8920 | 285.65 |
| 0.5826 | 300.04 | | |

4-Phenylbutan-2-one (1) + hexadecane(2)

| | | | |
|---|---|---|---|
| 0.2459 | 290.22 | 0.5920 | 307.04 |
| 0.2521 | 290.85 | 0.6483 | 306.99 |
| 0.2903 | 295.28 | 0.6709 | 306.96 |
| 0.3035 | 296.21 | 0.7028 | 306.71 |
| 0.3438 | 299.09 | 0.7288 | 306.53 |
| 0.3533 | 299.76 | 0.7457 | 306.40 |
| 0.3862 | 301.77 | 0.7760 | 306.00 |
| 0.4053 | 302.87 | 0.7991 | 305.26 |
| 0.4257 | 303.60 | 0.8161 | 304.38 |
| 0.4397 | 304.12 | 0.8521 | 302.01 |
| 0.4676 | 305.72 | 0.8714 | 299.27 |
| 0.5174 | 306.37 | 0.8981 | 294.54 |
| 0.5433 | 306.76 | | |

Benzyl ethanoate(1) + tetradecane(2)

| | | | |
|---|---|---|---|
| 0.3480 | 280.04 | 0.6619 | 288.22 |
| 0.3762 | 281.85 | 0.6909 | 288.18 |
| 0.4014 | 283.19 | 0.7069 | 288.12 |

Table 2 (continued)

| | | | |
|---|---|---|---|
| 0.4429 | 285.08 | 0.7088 | 288.12 |
| 0.4452 | 285.13 | 0.7198 | 288.08 |
| 0.4756 | 286.18 | 0.7213 | 288.05 |
| 0.5098 | 287.03 | 0.7619 | 287.65 |
| 0.5244 | 287.30 | 0.7887 | 287.11 |
| 0.5582 | 287.80 | 0.7958 | 286.91 |
| 0.5780 | 287.89 | 0.8243 | 285.61 |
| 0.5898 | 288.02 | 0.8476 | 283.23 |
| 0.6211 | 288.13 | 0.8721 | 280.17 |
| 0.6501 | 288.31 | | |
| Benzyl ethanoate(1) + hexadecane(2) | | | |
| 0.3915 | 287.83 | 0.6463 | 295.13 |
| 0.4011 | 288.15 | 0.6695 | 295.14 |
| 0.4319 | 289.36 | 0.6952 | 295.14 |
| 0.4512 | 290.50 | 0.7127 | 295.04 |
| 0.4616 | 290.85 | 0.7487 | 294.97 |
| 0.4634 | 290.92 | 0.7671 | 294.88 |
| 0.5110 | 292.85 | 0.8052 | 294.32 |
| 0.5289 | 293.43 | 0.8184 | 293.64 |
| 0.5609 | 294.04 | 0.8390 | 293.16 |
| 0.5745 | 294.34 | 0.8676 | 291.19 |
| 0.6006 | 294.70 | 0.8917 | 288.13 |
| 0.6235 | 294.98 | 0.9109 | 285.35 |

[a] standard uncertainties are: $u(x_1) = 0.0005$; $u(p) = 1$ kPa; the combined expanded uncertainty (0.95 level of confidence) for temperature is $U_c(T) = 0.1$ K

**Table 3**

Coefficients in eq. (1) for the Fitting of the ($x_1$, $T$) Pairs Given in Table 2 for Aromatic Polar Compound(1) + $n$-Alkane(2) Mixtures; $\sigma(T)$ is the Standard Deviation Defined by eq. (5).

| $N$ [a] | $m$ | $k$ | $\alpha$ | $T_c$/K | $x_{1c}$ | $\sigma(T)$/K |
|---|---|---|---|---|---|---|
| | | | 4-Phenylbutan-2-one(1) + dodecane(2) | | | |
| 26 | 3.792 | −949 | 0.906 | 292.3 | 0.552 | 0.05 |
| | | | | (296.4)[b] | (0.563)[b] | |
| | | | 4-Phenylbutan-2-one(1) + tetradecane(2) | | | |
| 27 | 2.899 | −333 | 0.723 | 300.2 | 0.593 | 0.05 |
| | | | | (304.2)[b] | (0.613)[b] | |
| | | | 4-Phenylbutan-2-one(1) + hexadecane(2) | | | |
| 25 | 3.488 | −627 | 0.532 | 306.9 | 0.653 | 0.05 |
| | | | | (310.2)[b] | (0.658)[b] | |
| | | | Benzyl ethanoate(1) + tetradecane(2) | | | |
| 25 | 3.372 | −589 | 0.445 | 288.2 | 0.668 | 0.06 |
| | | | | (291.3)[b] | (0.656)[b] | |
| | | | Benzyl ethanoate(1) + hexadecane(2) | | | |
| 24 | 2.728 | −223 | 0.476 | 295.2 | 0.698 | 0.04 |
| | | | | (298.3)[b] | (0.702[b] | |

[a] number of experimental data points; [b] DISQUAC result using interchange coefficients listed in Table 4

**Table 4**

**Dispersive (DIS) and Quasichemical (QUAC) Interchange Coefficients, $C_{ak,l}^{DIS}$ and $C_{sk,l}^{QUAC}$ ($l$ = 1, Gibbs energy; $l$ = 2, Enthalpy; $l$ = 3, Heat Capacity) for (a,k) Contacts[a] in Aromatic Polar Compound + $n$-Alkane Mixtures.**

| $C_nH_{2n+2}$ | $C_{ak,1}^{DIS}$ | $C_{ak,2}^{DIS}$ | $C_{ak,3}^{DIS}$ | $C_{ak,1}^{QUAC}$ | $C_{ak,2}^{QUAC}$ | $C_{ak,3}^{QUAC}$ |
|---|---|---|---|---|---|---|
| 4-phenylbutan-2-one | | | | | | |
| $n = 10$ | 2.33 | 1.5 | 2 | 5.75 | 5.75 | 2 |
| $n = 12$ | 2.17 | 1.5 | 2 | 5.75 | 5.75 | 2 |
| $n = 14$ | 2.07 | 1.5 | 2 | 5.75 | 5.75 | 2 |
| $n = 16$ | 1.98 | 1.5 | 2 | 5.75 | 5.75 | 2 |
| Benzyl ethanoate | | | | | | |
| $n < 10$ | −1.65 | 0.21 | | 2.8 | 2.25 | |
| $n \geq 10$ | −2 | 0.21 | | 2.8 | 2.25 | |

[a]type a, $CH_3$, $CH_2$ in $n$-alkanes, or in aromatic polar compounds considered; type k, CO in 4-phenylbutan-2-one, or COO in benzyl ethanoate

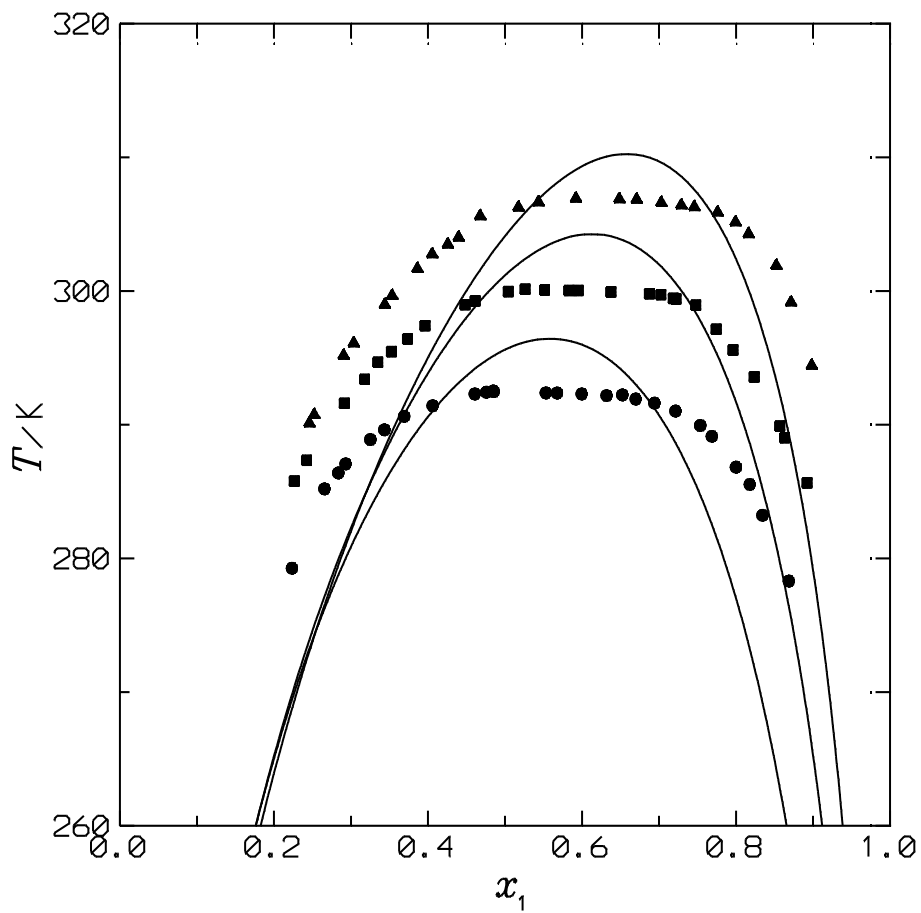

Figure 1. LLE for 4-phenylbutna-2-one(1) + *n*-alkane(2) mixtures. Points, experimental results (this work): (●), dodecane; (■), tetradecane (▲), hexadecane. Solid lines, DISQUAC calculations with interaction parameters listed in Table 4.

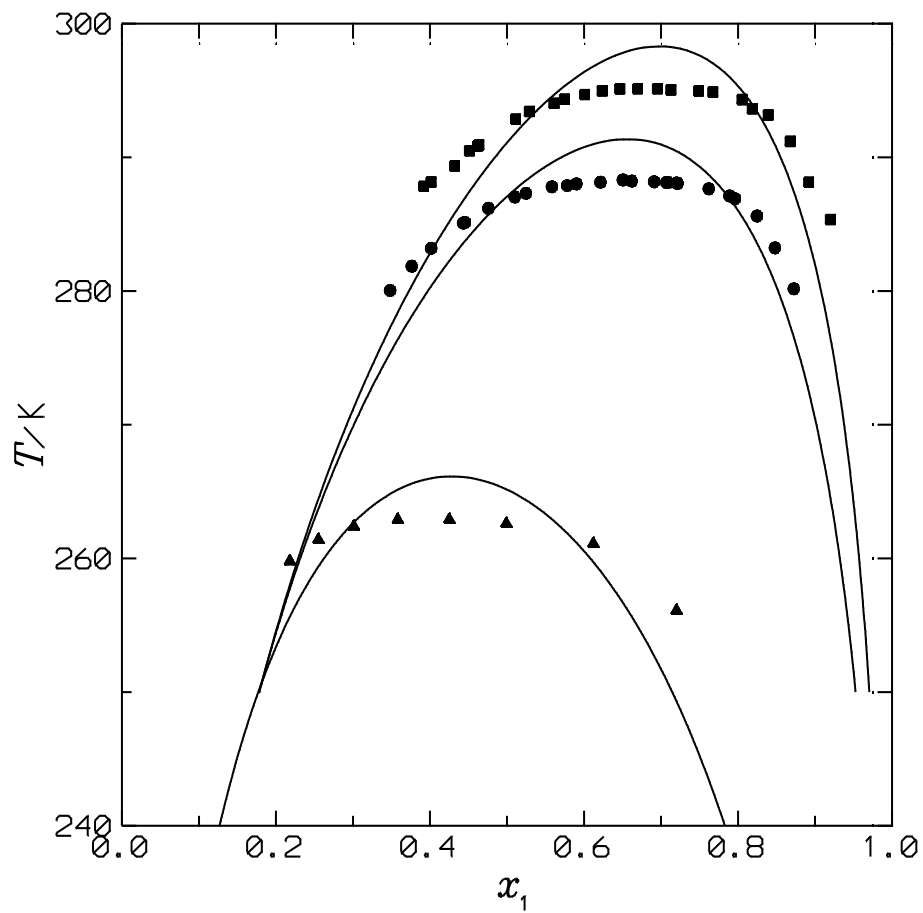

Figure 2. LLE for benzyl ethanoate(1) + *n*-alkane(2) mixtures. Points, experimental results: (▲), heptane;[41] (●), tetradecane; (■), hexadecane (this work). Solid lines, DISQUAC calculations with interaction parameters listed in Table 4.

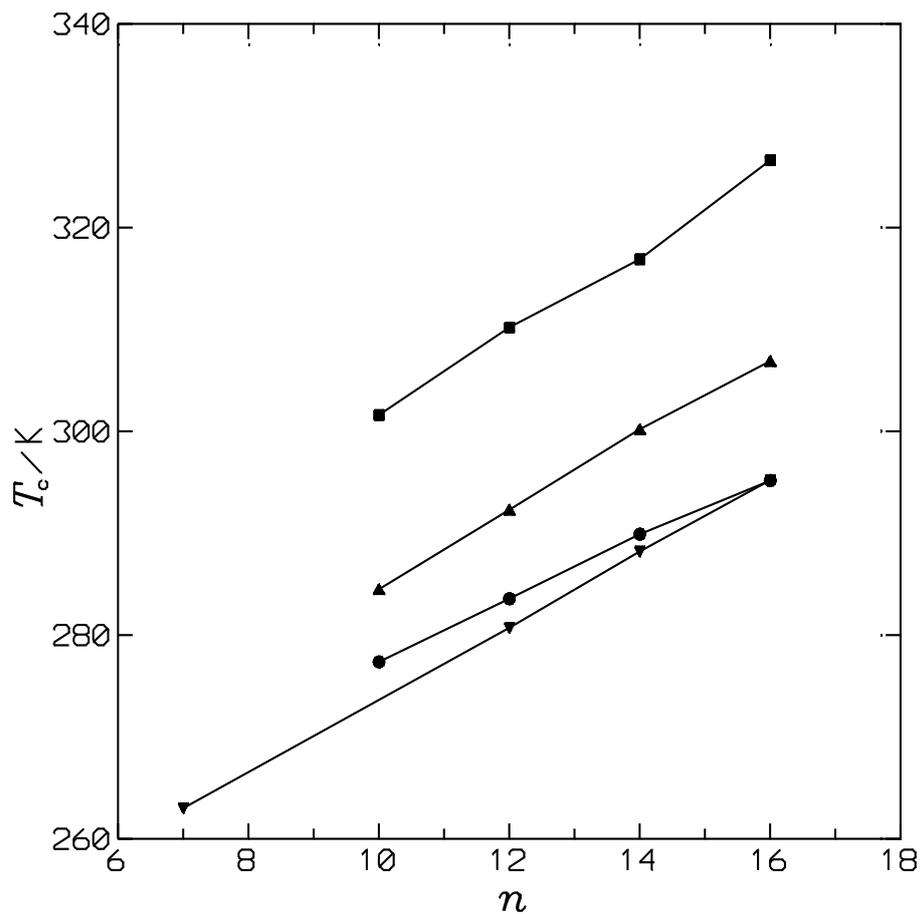

Figure 3.  Upper critical solution temperatures, $T_c$, vs. $n$, the numer of C atoms in the $n$-alkane for aromatic polar compound + $n$-alkane mixtures: (●), 1-phenylethanone;[11] (▲), 4-phenylbutan-2-one (this work);[13] (■), 1-phenylpropan-2-one;[13] (▼), benzyl ethanoate (this work).[41]

FOR TABLE OF CONTENTS USE ONLY

# LIQUID-LIQUID EQUILIBRIA FOR SYSTEMS CONTAINING 4-PHENYLBUTAN-2-ONE OR BENZYL ETHANOATE AND SELECTED ALKANES

CRISTINA ALONSO TRISTÁN[(1)], JUAN ANTONIO GONZÁLEZ*[(2)], FERNANDO HEVIA,[(2)], ISAÍAS GARCÍA DE LA FUENTE[(2)] AND JOSÉ CARLOS COBOS[(2)]

[(1)] Dpto. Ingeniería Electromecánica. Escuela Politécnica Superior. Avda. Cantabria s/n. 09006 Burgos, (Spain)

[(2)] G.E.T.E.F., Departamento de Física Aplicada, Facultad de Ciencias, Universidad de Valladolid, Paseo de Belén, 7, 47011 Valladolid, Spain,

*e-mail: jagl@termo.uva.es; Fax: +34-983-423136; Tel: +34-983-423757

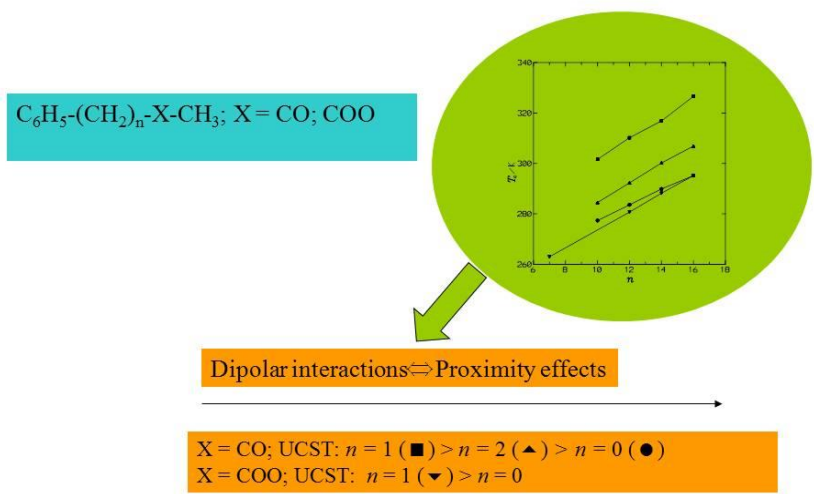

$C_6H_5$-$(CH_2)_n$-X-$CH_3$; X = CO; COO

Dipolar interactions ⇒ Proximity effects

X = CO; UCST: $n = 1$ (■) > $n = 2$ (▲) > $n = 0$ (●)
X = COO; UCST: $n = 1$ (▼) > $n = 0$